# Application of light pressure in the rheology of biopolymers

Sasun G. Gevorkian[a,d], Artashes V. Karmenyan[b,c,*], Chin-Kun Hu[a],

a. *Institute of Physics, Academia Sinica - Nankang, Taipei, Taiwan*
b. *Biophotonics and Molecular Imaging Research Center,*
  *National Yang Ming University, Taipei, Taiwan*
c. *Department of Physics, National Dong Hwa University, Hualien, Taiwan*
d. *Yerevan Physics Institute, Yerevan, Armenia*
  *[*]* artashes@mail.ndhu.edu.tw

**Abstract**
Precise measuring the deforming force and observation of the deformation dynamics along with difficulties of micro sample preparation are the main problems which make the micro-scale experiments difficult. Here we present a method for direct observation and measurements of mechanical parameters of elastic microfibrils in air using the pressure of optical light. For DNA microfibers, observation of the behavior dynamics of mechanical characteristics and the method sensitivity to changing the structure are demonstrated.

**Introduction**

Mechanical properties can provide insightful information not only on internal elastic forces, but also on the processes related to molecular movement involving stress relaxation of mechanical tensions in solid samples. Often this makes mechanical methods irreplaceable for investigating dynamics of complex solid state macromolecular systems [1]. For materials more complex than simple springs, where a spring constant relates force to elongation, the goal of mechanics is to provide quantitative parameters that define how a material will deform as a function of force, time and spatial orientation [2, 3]. Precise measuring of the deforming force value and its dynamics, controlling the frequency of the force are the main problems along with difficulties of micro sample preparation make the micro-scale experiments difficult. These problems particularly occur when investigating solid-state biopolymers [4-6]. In this case instability of the sample and scale of deforming force almost of the same order. Sample preparation and immobilizing techniques become more difficult due to the brittleness of solid biopolymers.

Many biological materials have hierarchic structure [7-9]. They are constructed from nano- to macro- level and often have self-organized structure. Many biological materials have the ability to react and adapt to environmental changes. This is one of the many reasons why investigation of supramolecular structures in biology is so important [10]. This hierarchic level is a critical link between molecular, microscopic, and macroscopic levels of living matter (examples are DNA and collagen microfibrils, amyloid microfibrils etc.). This is the level on which occur the properties that are difficult to observe on molecular level and "blurred" on macro-level [11, 12]. Biopolymers function in water environment, and methods that allow modifying that environment are important for investigating biopolymer-water interactions. Such methods generally deal with solid samples (protein and amino-acid micro crystals, DNA and collagen microfibrils). Between experiments on macroscopic samples of DNA and on a single molecule, there is a hierarchic level which will help to understand the connection between the former two. Such samples have diameter less than 1 μm. For DNA, it's the size of a chromatin (Fig. 1).

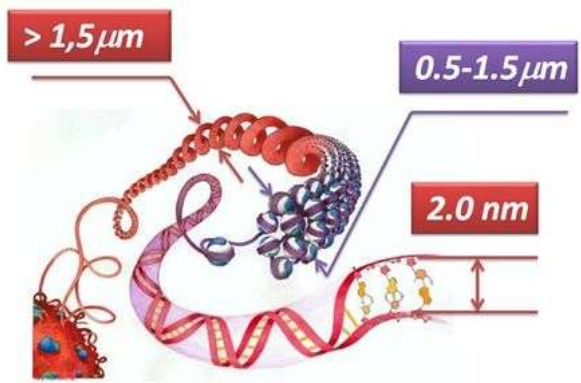

**Figure 1**. Hierarchy of DNA from molecule to chromatin. Our method operates with two levels above the molecular. Typical structure sizes are marked. DNA double spiral with size of 2 nm and chromatin with size tenth of μm. (Image adapted from watercolour by N. Bouvier for G. Almouzni).

After the first demonstration of the possibility of using light pressure for acceleration and trapping of particles [13], optical tweezers have been successfully used for decades for manipulation of biological cells, piconewton force measurements, investigation of single molecule, etc. [14-17].

In this present work, the method of measurements of fibrils' mechanical parameters is developed utilizing radiation pressure on microscopic dielectric beads (0.9-6 μm in diameter) attached to solid fiber sample. The method was tested on DNA microfibrils with diameter less than 1 μm with attached quartz beads (1.5 μm in diameter). It was shown that force of radiation pressure (around 80 pN) oscillates DNA microfibrils with both console and two-side fixed.

Sensitivity of the method to intramolecular changes in microsample was shown for the B-A transition of DNA caused by hydration change.

The introduced method allows "scanning" of a sample with frequency resolution less than 1 Hz. The method is readily available due to its simplicity.

The main principle of the suggested method is schematically shown in Fig. 2. For simplicity, the sample with both ends fixed is only shown. The sample with length $l$ (in our case it is DNA microfibril) is fixed between two quartz cylinders (Fig. 3). Radiation pressure is applied to the attached dielectric bead. Under external force the fibril undergoes deformation. Amplitude of initial deformation, $u(x)$, depends on external force, $p$, Young's modulus of the fibril, $E$, and coordinate, $x$. The main inertia momentum, $I$, characterizes the shape of sample cross section. For round cross section and diameter, $d$, we have $I = I_{min} = \frac{\pi d^4}{64}$

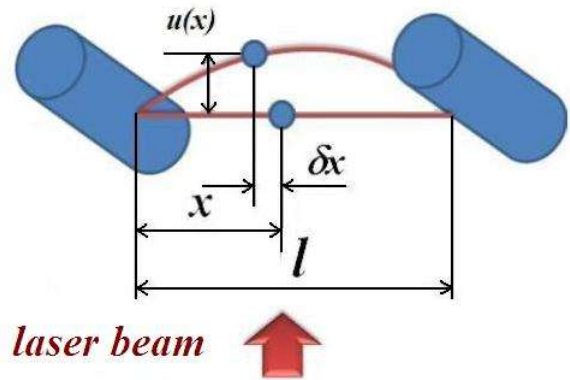

**Figure 2.** Schematic representation of the experiment. Fiber of DNA is stretched between two quartz cylinders. A quartz bead is shown on the fiber. Light pressure is periodically acting on this bead. Two fiber positions, as shown, are in equilibrium and deviation from equilibrium.

During the experiment, a fixed force was applied with variable interruption frequency to measure the oscillation amplitude and geometrical parameters of the sample. This approach allowed within a simple mechanical experiment to measure viscoelastic properties of the sample and make conclusions about internal processes.

A. For a sample fixed on both sides we have,

$$u(x) = \frac{p}{24EIl}(x^4 - 2lx^3 + l^2x^2),$$

and for maximum amplitude

$$y_{max} = \frac{pl^3}{384EI} \text{ when } x \to l/2$$

**B**. For console fixed sample,

$$u(x) = \frac{p}{24EIl}(x^4 - 2lx^3 + 6l^2x^2)$$

and $y_{max} = \frac{pl^3}{8EI}$ when $x \to l$

By definition logarithmic decrement of dampening is directly proportional to natural logarithm of the ratio of two amplitudes $x_0$ and $x_N$ with $N$ denotes the number of periods and $\delta = \frac{1}{N} ln \frac{x_0}{x_N}$

And the quality of oscillating system is defined by $Q = \frac{\pi}{\delta}$

(For formula derivation, please refer to **Supplementary 1**).

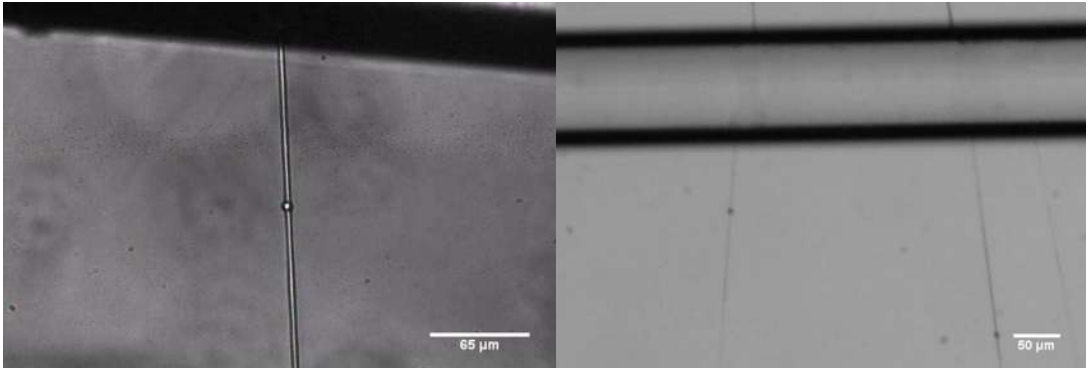

**Figure 3**. Pictures of different DNA samples with the beads.

In the preparation of the oriented microfibrils from saturated DNA solution, surface tension forces of the newly formed filaments orient DNA molecules along the long axis [20]. For perfect orientation of long macromolecules during the drying process, it is important to have an additional opportunity to orient the molecules in the saturated solution. Known methods are using the varying drawing speed [20]. We swing the samples using light pressure not only just during the measurements but also during the preparation of samples in the drying process. The shape and form of the samples obtained through this process are close to ideal cylinder and this is very important when calculating and interpretation of the data about the mechanical properties of the materials [1, 2]. Fig. 4 shows the changing of amplitude of oscillation of fiber in the process of drying of the sample, when relative humidity (RH) changes from 95 to 85%. In this range of RH, DNA has B-form [e.g. 5, 18]. Decreasing main amplitude from 120 to 110 relative units indicates that the Young's modulus (stiffness) of microfibrils increases. The amplitude of the periodic oscillations superimposed on the basic curve characterizes the internal friction [1].

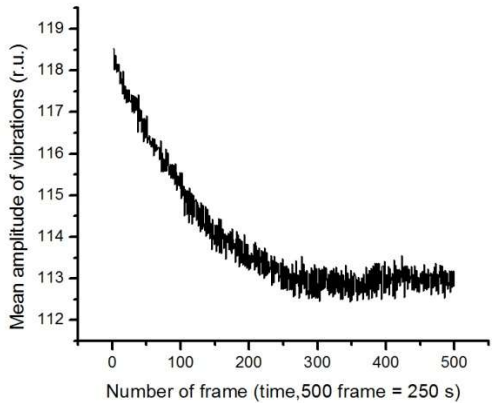

**Figure 4.** Changing the amplitude of oscillation of fiber on drying. Initial RH in the chamber is 95%. The final RH is 85%. DNA is found in B-form.

The amplitude of these vibrations increases during drying, which demonstrates the decrease of internal friction in the samples.

This phenomenon is a typical behavior of most polymers [1], previously observed for larger films and for fibrils of DNA studied by other mechanical methods [5].

The drying process in microfibrils of DNA in the A-form is compared in Fig. 5. In this case the RH changes from 60 to 50%. Moreover, a smooth increase in stiffness (large amplitude decreases) can also be observed in the background while the internal friction is not changing (since the amplitude of the periodic oscillations does not change).

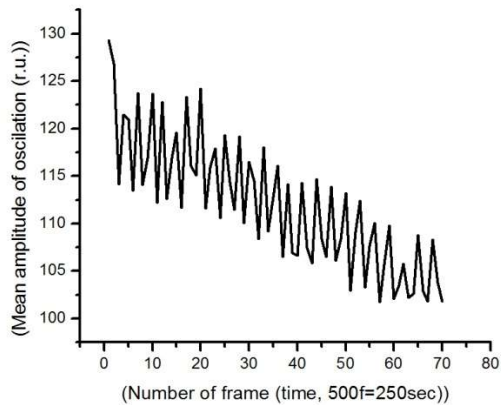

**Figure 5.** Changing the amplitude of oscillation of fiber on drying. Initial RH in the chamber is 60%. The final RH is 50%. DNA is found in A-form.

From our result it is observed that Young's modulus decreases with drying as the average amplitude drops from 120 to 105 units for A-form of the DNA fibril. Meanwhile the amplitude of low frequency oscillations remains constant. It is known [18] that internal structural changes within DNA molecules are completed under these values of humidity.

DNA conformations and transition between them are basics of important biological processes. In important processes such as replication, a DNA reparation cannot be completely understood without considering physical properties of conformations and transitional dynamics [19]. Despite the fact that in water solutions DNA is mostly in B-conformation, it is often observed that DNA undergoes conformational changes. Though conformational transition was discovered more than 50 years ago it is still a subject of research. During this transition DNA undergoes conformational [5] (changes the pitch of the spiral) and physical changes. The transition itself is caused by change of hydration level [5, 20]. We decided to check our method's sensitivity to this transition. Investigation of mechanical properties of DNA amorphous films and oriented fibrils were performed previously by micromechanical methods [5]. In this work Gevorkian *et al*. noticed non homogeneous character of B-A transition caused by change of RH.

Sample was prepared (see Materials and Methods) and was kept under temperature t = 25$^o$C and RH = 95%. The process was observed by registering amplitude of oscillations at constant frequency and oscillating force. After amplitude was stabilized for several hours, humidity gradient was created in the chamber. RH was decreased from 95 to 35%. In order to create the gradient airspace, the hermetic chamber was brought into contact with a vessel with RH = 32%. In Fig. 6 the typical data obtained in such experiment are shown. In the beginning (RH=95%) DNA is in B-form, while at the end (RH=32%) it is in A-form. The amplitude was changed during the phase transition of DNA conformation. The reverse transition was also investigated. For this the RH was allow stabilized at 32% before it was reversed back to 95%. The data is shown in Fig. 7. This article is mainly about the method itself and hence the authors have decided to publish the detailed description and experimental data about the physical properties of DNA in future. Here we mention the sensitivity of the method to conformational transition of DNA. It should also be noticed that physical condition of DNA between B- and A- conformations, which was described in [5], is further confirmed by our experiments. Apart from known two conformations that differ by the pitch of the spiral, there is a third conformation which has different elasticity characteristics (i.e. more rigid). There is a hysteresis between two transitions, which is also an interesting subject for understanding conformational physics.

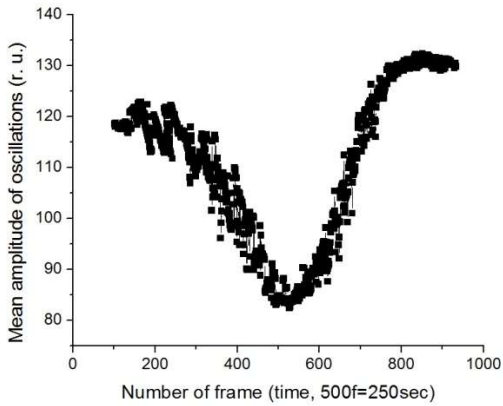

**Figure 6.** B-A transition. The amplitude changes during the phase transition of DNA conformation. Initial RH in the chamber is 95%. The final RH is 35%. At the beginning of experiments we have B-form DNA, and at the end an A-form. A new state can be clearly seen between the two conformations.

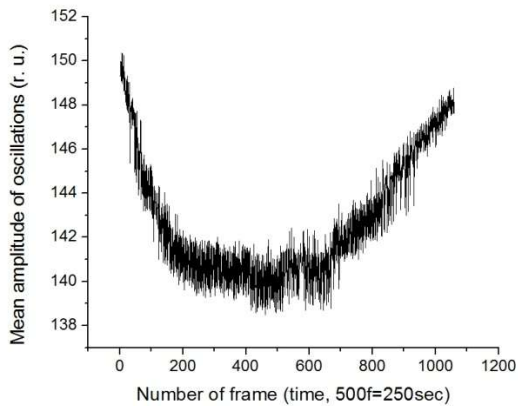

**Figure 7.** A-B transition. The amplitude change during the phase transition of DNA conformation. Initial RH in the chamber is 35%. The final RH is 95%. At the beginning of experiments we have A-form DNA, in the end a B-form.

Investigation of frequency dependence of viscoelastic properties of polymers is an essential attribute of mechanics research. It is known that many important properties of biopolymers were discovered by such research [5]. Mechanical response of the system on external force depends on frequency of that force. By changing that frequency we observe the object through different frequency "windows". For biological objects low frequencies (lower than 100 Hz) are of particular importance. In this frequency range most of the processes in living organisms occur. Examining physical properties of biologically important molecules in water solution at those frequencies is a complicated (almost impossible) task. All vibrations are dampened as inharmonic oscillator.

Our method allows smooth changing of stimulation frequency below 100 Hz and gives a possibility to "scan" the object in the physiologic range of frequencies. Fig. 8 displays the result of a typical experiment. Relative humidity was kept constant at RH = 90% and the stimulating frequency was variable throughout the experiment. It was shown that frequency increment lead to "releasing" of otherwise restrained movements. It was expressed by increasing elasticity of the sample (reducing amplitude) and abrupt changing of low frequency oscillations. The thorough analysis of those processes related to DNA physics is a responsible task and is left for future investigation.

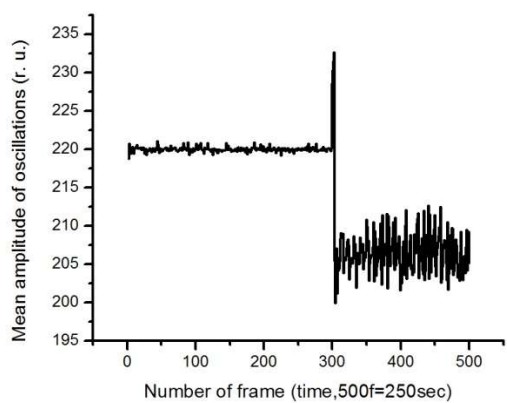

**Figure 8.** RH and temperature in the chamber are fixed at 90% and 25$^{o}$C, respectively. The frequency of the disturbing force is changed from 4.3 to 8.7 Hz.

Here, we have successfully demonstrated a new method of solid DNA fibrils examination. It is for the first time radiation pressure has been utilized as external oscillating force. This method allows examining of both console fixed samples and samples immobilized on two ends. Further, it also allows to examine the relation of mechanical properties to parameters such as humidity (sample hydration), temperature (in physiological area of values), sample size and frequency. The sensitivity to conformational changes of DNA was also shown.

**Materials and Methods**

**Preparation of oriented DNA fibrils.**

All methods of oriented DNA fibril preparation adopt the following procedure. Fibril strain is performed in saturated water solution. The strain rate must be sufficient for long molecules to relocate along the axis of the sample. By controlling the drying rate, the rate of molecular orientation can be controlled. Predominant orientation along the axis is provided by surface tension and the straining force itself. The value of the force is set by various methods [20, 21]. This is the way of creating samples for micromechanical methods [5] and X-ray crystallography. Diameter of samples produced with this method varies from μm to mm.

In present work, the proposed method is as follows. A sample of dry DNA from calf thymus (Sigma Aldrich D1501 Deoxyribonucleic acid sodium salt from calf thymus) (2.5 mg) was allowed to swell in 1 ml distilled and deionized water for several hours. This ratio allowed DNA to swell without being dissolved. After allowing it to swell for at least 3 hours, distilled water was gradually added (by droplets on the tip of a glass stick) and the solution was mixed until it reached the condition of homogeneous dense solution. The process could be visually controlled under binocular microscope. Separately silica microspheres solution (from 0.9 to 6 μm in diameter, Bangs Laboratories Inc.) in distilled water was prepared. Suspension with dielectric beads was mechanically mixed until homogeneous condition was reached. No bead conglomerates were allowed. A small portion (1-2 μl) of the suspension was mixed with DNA solution under vibration on vortex mixer (Grant-bio PV-1, Grant Instruments Ltd). Two quartz cylinders with diameter up to 125 μm (clad from silica fiber, 460HP, Thorlabs) were fixed on cover glass with double sided sticky tape. The mixed solution of DNA and the particles were dropped on the coverglass and DNA fibrils were strained between two quartz cylinders. Manually we were able to strain the fibrils with diameter of 0.29 μm. However, it is also possible to utilize micro-manipulators. It allows producing of fibrils with even cross section along the sample. From about 10 strained fibrils a few suitable ones with one or few attached beads were chosen for the experiment. In Fig. 3 one of the manually prepared samples is displayed. For preparing a console fixed sample, a regular strained fibril was bisected by tightly focused laser beam.

To prevent the influence of environment on the samples and the humidity variations directly in the area of the fibril the sample was closed with cover glass, which was fixed by the perimeter.

For performing of the experiments with varying humidity one side of the sample was open and wick had been introduced inside the sample to supply with CaCl$_2$ solution with different concentrations [22].

For preparing samples (Fig. 3) with even cross section and maximum molecular orientation two important parameters were controlled *viz.* level of relative humidity and vibration during the drying of sample. In the drying chamber necessary RH level was created with solutions of CaCl$_2$. At the same time samples were oscillated by radiation pressure. Fibril was strained from saturated solution which indicated 100% initial RH in the chamber which was dropped to 85% later.

**Experimental setup.**

For the force-oscillation and displacement measurements invert microscope (Olympus IX-71) was used to influence and image the DNA fiber – sample (DNA, polymers), equipped with 40x and 20x microscope objectives (Olympus UPlanFl 40x/0.75 and UPlanApo 20x/0.70). Images were taken with monochrom Camera WAT-120N (Watec) connected with PC through the National Instruments PC card. Frames were recorded with homemade software on LabView (National Instruments Inc.) allowing to record up to 10 frame per second. For high repetition rate and possibility to select region of interest CMOS camera Lu175C (Lumenera), connected to USB port, was used.

Titanium:Sapphire lasers (Spectra Physics) at 800 nm wavelength was used as a source of light pressure and allowed using average power up to 270 mW after microscope objective. This power is much higher than necessary in the experiments. Repetition rate of light beam oscillations was set and adjusted with optical chopper (300CD; Scitec Instruments) placed on the entrance of beam to the microscope. Variable neutral density filter (NDC-100C-4M, Thorlabs) was utilized before beam input in microscope to control and change the intensity of light. Precise position of the sample on XY plane was controlled and fixed with motorized stage MR-IX (Tanlian). Vertically positioned light beam was driven to the beads on the fiber with focus controller ES10ZE (Prior Scientific). Power meter, Laser-Star Dual channel (Ophire) was used for average power measurements.

The measurements were done as follow.

First and foremost, the coordinates of the point of a laser irradiation had to be precisely fixed before positioning the sample on the microscope stage.

After positioning the sample on the microscope stage, with closed laser irradiation, the fiber was selected for the measurements, the bead attached on the fiber was moved to the point of irradiation, the selected/necessary repetition rate of an optical chopper was adjusted, and then the laser beam was opened to irradiate the sample. The bead position was accurately adjusted according to the laser beam position to illuminate the center of the bead.

The power of irradiation which allowed reliable detection of the sample oscillation without any damaging or deforming of the sample during measurements was selected using filters.

For the data acquisition the recording time was set up to 40 min. To observe the transitions between the A and B form/state of DNA, the recording was performed until full visual stabilization of the sample.

The obtained images were processed using standard software ImageJ (tracking plugins), Origin 6.0, MATLAB (R2012).

# Supplementary 1

The task of calculating the parameters of externally forced oscillation of hard fixed homogenous string with length, $0 < l < x$, and density, $\rho$, is reduced to solution of the following equation

$$\frac{\partial^2 u}{\partial t^2} = a^2 \frac{\partial^2 u}{\partial x^2} + g(x,t), \qquad (1)$$

with the following initial conditions

$$u|_{t=0} = u_0(x), \qquad \frac{\partial y}{\partial x}\bigg|_{t=0} = u_1(x), \qquad (2)$$

and boundary conditions

$$u|_{x=0} = 0, \qquad u|_{x=l} = 0 \qquad (3)$$

Explanation: $g = \frac{p}{\rho}$, where $\rho$ is the linear density of the string. The solution is sought in the following form, $u = v + w$, where:

1) $v$ is the solution of equation (1) with boundary conditions (3) and initial conditions

$$u|_{t=0} = 0, \qquad \frac{\partial u}{\partial t}\bigg|_{t=0} = 0$$

(Solution of heterogeneous equation with both zero initial and terminal conditions)

2) $w$ is the solution of homogeneous equation $\frac{\partial^2 u}{\partial t^2} = a^2 \frac{\partial^2 u}{\partial x^2}$, with boundary conditions (3) and initial conditions (2) (solution of homogeneous equation).

**1.** In order to find $w$ we need to solve the equation $\frac{\partial^2 w}{\partial t^2} = a^2 \frac{\partial^2 w}{\partial x^2}$

with boundary conditions $w|_{x=0} = 0, \quad w|_{x=l} = 0,$

and initial conditions $w|_{t=0} = u_0(x), \quad \frac{\partial w}{\partial t}\bigg|_{t=0} = u_1(x)$

Formulation of the task completely matches with hyperbolic-type equation which is solved in the first chapter of the article using Fourier's method. The solution is the following function:

$$w(x,t) = \sum_{n=1}^{\infty} w_n(x,t) = \sum_{n=1}^{\infty} \left(C_n \cos\left(\frac{\pi n}{l} at\right) + D_n \sin\left(\frac{\pi n}{l} at\right)\right) \sin\left(\frac{\pi n}{l} x\right)$$

with coefficients

$$C_n = \frac{2}{l} \int_0^l u_0(\xi) \sin\left(\frac{\pi n}{l}\xi\right) d\xi$$

$$D_n = \frac{2}{\pi n a} \int_0^l u_1(\xi) \sin\left(\frac{\pi n}{l}\xi\right) d\xi$$

It should be mentioned that proper values of $\lambda = \left(\frac{\pi n}{l}\right)^2$ and $X_n(x) = \sin\left(\frac{\pi n}{l}x\right)$ were also found during solution.

**2.** In order to find v we have to solve heterogeneous equation

$$\frac{\partial^2 v}{\partial t^2} = a^2 \frac{\partial^2 v}{\partial x^2} + g(x,t),$$

with terminal conditions $v|_{x=0} = 0, \quad v|_{x=l} = 0,$

and initial conditions $v|_{t=0} = 0, \quad \frac{\partial v}{\partial t}|_{t=0} = 0$

The solution is the following:

$$v(x,t) = \sum_{n=1}^{\infty} T_n(t) \sin\left(\frac{\pi n}{l}x\right) \qquad (4)$$

Note: proper values and proper functions of previous equation were used.

Now we use equation (4) in the initial equation (1). $\frac{\partial^2 v}{\partial t^2} = a^2 \frac{\partial^2 v}{\partial x^2} + g(x,t)$

$$\frac{\partial^2}{\partial t^2}\left(\sum_{n=1}^{\infty} T_n(t) \sin\left(\frac{\pi n}{l}x\right)\right) = a^2 \frac{\partial^2}{\partial x^2}\left(\sum_{n=1}^{\infty} T_n(t) \sin\left(\frac{\pi n}{l}x\right)\right) + g(x,t)$$

$$\sum_{n=1}^{\infty} T_n''(t) \sin\left(\frac{\pi n}{l}x\right) = -\left(\frac{\pi n}{l}a\right)^2 \sum_{n=1}^{\infty} T_n(t) \sin\left(\frac{\pi n}{l}x\right) + g(x,t)$$

$$\sum_{n=1}^{\infty} T_n''(t) \sin\left(\frac{\pi n}{l}x\right) + \left(\frac{\pi n}{l}a\right)^2 \sum_{n=1}^{\infty} T_n(t) \sin\left(\frac{\pi n}{l}x\right) = g(x,t)$$

$$\sum_{n=1}^{\infty} \left(T_n''(t) + \left(\frac{\pi n}{l}\right)^2 T_n(t)\right) \sin\left(\frac{\pi n}{l}x\right) = g(x,t)$$

Let's expand the function (x, t) to Fourier series in the interval $0 < x < l$

$$g(x,t) = \sum_{n=1}^{\infty} g_n(t) \sin\left(\frac{\pi n}{l}x\right),$$

and compare last two equations

$$\sum_{n=1}^{\infty}\left(T_n''(t) + \left(\frac{\pi n}{l}\right)^2 T_n(t)\right) \sin\left(\frac{\pi n}{l}x\right) = \sum_{n=1}^{\infty} g_n(t) \sin\left(\frac{\pi n}{l}x\right)$$

From this we can compose differential equation:

$T_n''(t) + \left(\frac{\pi n}{l}a\right)^2 T_n(t) = g_n(t)$, where $g_n(t) = \frac{2}{l}\int_0^l g(\xi,t) \sin\left(\frac{\pi n}{l}\xi\right) d\xi$

The equation should be solved for initial conditions:

$$T_к(0) = 0, \quad T_k'(0) = 0, \quad (k = 1,2,3\ldots)$$

The solution can be presented in the following form:

$$T_n(t) = \frac{l}{n\pi a}\int_0^t g_n(\tau) \sin\left(\frac{\pi n a}{l}(t-\tau)\right) d\tau$$

If we use this in the equation (4) we'll have the expression for $v$.

**Conclusion:**

The solution of the initial task is following:

$$u(x,t) = \sum_{n=1}^{\infty} T_n(t) \sin\left(\frac{\pi n}{l}x\right) + \sum_{n=1}^{\infty}\left(C_n \cos\left(\frac{\pi n}{l}at\right) + D_n \sin\left(\frac{\pi n}{l}at\right)\right) \sin\left(\frac{\pi n}{l}x\right)$$

$$T_n(t) = \frac{l}{n\pi a}\int_0^t g_n(\tau) \sin\left(\frac{\pi n a}{l}(t-\tau)\right) d\tau$$

$$C_n = \frac{2}{l}\int_0^l u_0(\xi) \sin\left(\frac{\pi n}{l}\xi\right) d\xi \quad \frac{\partial^2 u}{\partial t^2} = 120a^2 \frac{\partial^2 u}{\partial x^2}$$

During the solution of initial task it was supposed that $a$ of the equations (1-3) is the tension of oscillating string and is constant and does not depend on excitation force, frequency and is homogeneous along the whole length.

**Task solution example: calculation of forced oscillations of the string**

Solve the task with the method of separating variables:

$u_{tt} = u_{xx} + 2b, \ (b = const, \ 0 < x < l), \ u\big|_{x=0} = u\big|_{x=l} = 0, \ u\big|_{t=0} = u_t\big|_{t=0} = 0$

**Solution:**

Solution of the task is sought as a sum, $u = v + w$,

where:

1) $v$ is the solution of heterogeneous equation $u_{tt} = u_{xx} + 2b$ with terminal conditions,

$$u\big|_{x=0} = u\big|_{x=l} = 0,$$

and initial conditions

$$u\big|_{t=0} = u_t\big|_{t=0} = 0$$

2) $w$ is the solution of homogeneous equation

$$\frac{\partial^2 u}{\partial t^2} = a^2 \frac{\partial^2 u}{\partial x^2}$$

with terminal conditions $u_{tt} = u_{xx}$,

and initial conditions $u\big|_{t=0} = u_t\big|_{t=0} = 0$

**1.** In order to find $w$, we have to solve the equation

$$u_{tt} = u_{xx}$$

with terminal conditions

$$w\big|_{x=0} = w\big|_{x=l} = 0,$$

and initial conditions $w\big|_{t=0} = 0, \ w_t\big|_{t=0} = 0$

Formulation of the task completely matches with hyperbolic-type equation which is solved in the first chapter of the article using Fourier's method with parameters:

$$a = 1, l = l, u_0 = 0, u_1 = 0$$

The solution is following:

$$w(x,t) = \sum_{n=1}^{\infty} u_n(x,t) = \sum_{n=1}^{\infty} \left(C_n \cos\left(\frac{\pi n}{l} at\right) + D_n \sin\left(\frac{\pi n}{l} at\right)\right) \sin\left(\frac{\pi n}{l} x\right)$$

with coefficients

$$C_n = \frac{2}{l} \int_0^l u_0(\xi) \sin\left(\frac{\pi n}{l}\xi\right) d\xi,$$

and

$$D_n = \frac{2}{\pi n a} \int_0^l u_1(\xi) \sin\left(\frac{\pi n}{l}\xi\right) d\xi$$

After placing known values:

$$w(x,t) = \sum_{n=1}^{\infty} \left(C_n \cos\left(\frac{\pi n}{l}t\right) + D_n \sin\left(\frac{\pi n}{l}t\right)\right) \sin\left(\frac{\pi n}{l}x\right)$$

with coefficients

$$C_n = \frac{2}{l} \int_0^l 0 \sin\left(\frac{\pi n}{l}\xi\right) d\xi = 0,$$

and

$$D_n = \frac{2}{\pi n} \int_0^l 0 \sin\left(\frac{\pi n}{l}\xi\right) d\xi = 0$$

Proper values

$$\lambda = \left(\frac{\pi n}{l}\right)^2,$$

and proper functions

$$X_n(x) = \sin\left(\frac{\pi n}{l}x\right)$$

As a result all coefficients are equal to 0, therefore

$$w(x,t) = 0$$

**2.** In order to find $v$, we have to solve the equation $v_{tt} = v_{xx} + 2b$ with terminal conditions

$$v\big|_{x=0} = 0, \quad v\big|_{x=l} = 0,$$

and initial conditions

$$v\big|_{t=0} = v_t\big|_{t=0} = 0$$

The solutions is

$$v(x,t) = \sum_{n=1}^{\infty} T_n(t) \sin\left(\frac{\pi n}{l} x\right)$$

Placing it to the initial equation $v_{tt} = v_{xx} + 2b$

$$\frac{\partial^2}{\partial t^2}\left(\sum_{n=1}^{\infty} T_n(t) \sin\left(\frac{\pi n}{l} x\right)\right) = \frac{\partial^2}{\partial x^2}\left(\sum_{n=1}^{\infty} T_n(t) \sin\left(\frac{\pi n}{l} x\right)\right) + 2b$$

$$\sum_{n=1}^{\infty} T_n''(t) \sin\left(\frac{\pi n}{l} x\right) = -\left(\frac{\pi n}{l}\right)^2 \sum_{n=1}^{\infty} T_n(t) \sin\left(\frac{\pi n}{l} x\right) + 2b$$

$$\sum_{n=1}^{\infty} T_n''(t) \sin\left(\frac{\pi n}{l} x\right) + \left(\frac{\pi n}{l}\right)^2 \sum_{n=1}^{\infty} T_n(t) \sin\left(\frac{\pi n}{l} x\right) = 2b$$

$$\sum_{n=1}^{\infty} \left( T_n''(t) + \left(\frac{\pi n}{l}\right)^2 T_n(t) \right) \sin\left(\frac{\pi n}{l} x\right) = 2b$$

According to common method we have to expand the $2b$ constant to Fourier series in the interval $0 < x < l$

$$2b = \sum_{n=1}^{\infty} g_n(t) \sin(nx)$$

$$g_n(t) = \frac{2}{l} \int_0^l 2b \sin\left(\frac{\pi n}{l} \xi\right) d\xi = \frac{4b}{l} \frac{l}{\pi n} \int_0^l \sin\left(\frac{\pi n}{l} \xi\right) d\frac{\pi n}{l} \xi = -\frac{4b}{\pi n} \cos\left(\frac{\pi n}{l} \xi\right)\bigg|_0^l =$$

$$= -\frac{4b}{\pi n}\left(\cos\left(\frac{\pi n}{l} l\right) - \cos\left(\frac{\pi 0}{l} \xi\right)\right) = -\frac{4b}{\pi n}(\cos(\pi n) - \cos(0)) = \frac{4b}{\pi n}(1 - (-1)^n)$$

After placing this expansion to the equation we'll have

$$\sum_{n=1}^{\infty} \left( T_n''(t) + \left(\frac{\pi n}{l}\right)^2 T_n(t) \right) \sin\left(\frac{\pi n}{l} x\right) = \sum_{n=1}^{\infty} \frac{4b}{\pi n}(1 - (-1)^n) \sin\left(\frac{\pi n}{l} x\right)$$

From here we can compose differential equation:

$$T_n''(t) + \left(\frac{\pi n}{l}\right)^2 T_n(t) = \frac{4b}{\pi n}(1 - (-1)^n)$$

And this equation can be solved for zero initial conditions.

$$T_k(0) = 0, \qquad T'_k = 0, \quad (k = 1,2,3 \ldots)$$

There is no need to manually solve this because the formula for solving such equations is mentioned in theoretical part:

$$T_n(t) = \frac{l}{n\pi a} \int_0^t g_n(\tau) \sin\left(\frac{\pi n a}{l}(t-\tau)\right) d\tau$$

$$T_n(t) = \frac{4b}{\pi n} \frac{l}{\pi n} (1$$

$$- (-1)^n) \int_0^t \sin\left(\frac{\pi n}{l}(t-\tau)\right) d\tau$$

$$= -\frac{4bl}{\pi^2 n^2} \frac{l}{\pi n} (1 - (-1)^n) \int_0^t \sin\left(\frac{\pi n}{l}(t-\tau)\right) d\frac{\pi n}{l}(t-\tau) =$$

$$= -\frac{4bl^2}{\pi^3 n^3} (1 - (-1)^n) \cos\left(\frac{\pi n}{l}(t-\tau)\right)\Big|_0^t = -\frac{4bl^2}{\pi^3 n^3} (1$$

$$- (-1)^n) \left(\cos\left(\frac{\pi n}{l}(t-t)\right) - \cos\left(\frac{\pi n}{l}(t-0)\right)\right)$$

$$= -\frac{4bl^2}{\pi^3 n^3} (1 - (-1)^n) \left(\cos(0) - \cos\left(\frac{\pi n}{l}t\right)\right)$$

Thus we'll have the equation for $v$

$$v(x,t) = \sum_{n=1}^\infty -\frac{4bl^2}{\pi^3 n^3} (1 - (-1)^n) \left(1 - \cos\left(\frac{\pi n}{l}t\right)\right) \sin\left(\frac{\pi n}{l}x\right)$$

$$= -\frac{4bl^2}{\pi^3} \sum_{n=1}^\infty \frac{(1-(-1)^n)}{n^3} \left(1 - \cos\left(\frac{\pi n}{l}t\right)\right) \sin\left(\frac{\pi n}{l}x\right) =$$

Note: for even value of $n$ multiplier $(1 - (-1)^n)$ becomes zero. As we are calculating solutions that are not equal to zero, then for the answer we'll take only non-zero summands. So for $n = 2k + 1$ (odd)

$$v(x,t) = -\frac{8bl^2}{\pi^3} \sum_{n=1}^\infty \frac{1}{(2k+1)^3} \left(1 - \cos\left(\frac{\pi(2k+1)}{l}t\right)\right) \sin\left(\frac{\pi(2k+1)}{l}x\right)$$

**Answer:**

Solution of the initial task:

$$u(x,t) = w(x,t) + v(x,t) = v(x,t)$$

$$= -\frac{8bl^2}{\pi^3} \sum_{n=1}^{\infty} \frac{1}{(2k+1)^3} \left(1 - \cos\left(\frac{\pi(2k+1)}{l}t\right)\right) \sin\left(\frac{\pi(2k+1)}{l}x\right)$$

In case sinusoidal oscillations by using Young's modulus, $E$, and main inertia momentum we'll have:

**A**. For a string with both ends fixed

$u(x) = \frac{p}{24EIl}(x^4 - 2lx^3 + l^2x^2)$ and for maximum amplitude we have $y_{max} = \frac{pl^3}{384EI}$, when $x \to {}^l/_2$

**B**. For console-fixed string

$u(x) = \frac{p}{24EIl}(x^4 - 2lx^3 + 6l^2x^2)$   $y_{max} = \frac{pl^3}{8EI}$, when $x \to l$ for a string with round cross section with diameter $d$ we have

$$I = I_{min} = \frac{\pi d^4}{64}$$